\documentclass[twocolumn,showpacs,amssymb,aps,nofootinbib,floatfix]{revtex4-1}

\usepackage{epsfig}
\newcommand{\eqcomma}{,\phantom{AA}}
\newcommand{\order}[1]{ \mathcal{O} \left( #1 \right) }
\newcommand{\ave}[1]{\left\langle #1 \right\rangle}
\newcommand{\mod}[1]{\left| #1 \right|}
\newcommand{\refeq}[1]{Eq. (\ref{#1})}

\begin{document} 
\hbadness=10000

\title{Longitudinal hydrodynamics from event-by-event Landau initial conditions }

\author{Abhisek Sen$^a$,Jochen Gerhard$^b$,Giorgio Torrieri$^{c}$,Kenneth Read$^{da}$,Cheuk-Yin Wong$^d$}
\affiliation{$^a$University of Tennessee, Knoxville, TN  37996}
\affiliation{$^b$Frankfurt Institute for Advanced Studies (FIAS), Frankfurt am Main, Germany}
\affiliation{$^c$ IFGW, Universidade Estadual de Campinas, Campinas, S$\tilde{a}$o Paulo, Brazil}
\affiliation{$^d$Oak Ridge National Laboratory, Oak Ridge, TN  37831}

\date{\today}

\begin{abstract}
We investigate three-dimensional ideal hydrodynamic evolution, with Landau initial conditions, incorporating event-by-event variation with many events and transverse density inhomogeneities.
We show that the transition to boost-invariant flow occurs too late for realistic setups, with corrections of $\order{20-30\%}$ expected at freezeout for most scenarios.
Moreover, the deviation from boost-invariance is correlated with both transverse flow and elliptic flow, with the more highly transversely flowing regions also showing the most violation of boost invariance.    Therefore, if longitudinal flow is not fully developed at the early stages of heavy ion collisions, hydrodynamics where boost-invariance holds at mid-rapidity is inadequate to extract transport coefficients of the quark-gluon plasma.    We conclude by arguing that developing experimental probes of boost invariance is necessary, and suggest some promising directions in this regard.
\end{abstract}


\maketitle
The quantitative modeling of matter produced in high energy heavy ion collisions with relativistic hydrodynamics is now a well-established field, following the widely cited announcement that matter produced at the relativistic heavy ion collider (RHIC), 
behaves as a ``perfect fluid''
\cite{whitebrahms,whitephobos,whitestar,whitephenix,sqgpmiklos,sqgpshuryak}.
The evidence for this behavior comes from the successful
modeling of RHIC anisotropic flow by boost-invariant hydrodynamics
\cite{tauv2,shuryak,romatschke,huovinen,chojnacki,hirano1,hirano2}. 
It is now clear that the same fluid-like behavior persists at the LHC \cite{alice,cms,atlas}.
It is commonly argued that, given precise enough data on soft physics, chiefly momentum spectra and their azimuthal anisotropy, the transport coefficients of matter created in ultrarelativistic heavy ion collisions can be quantitatively constrained.  Several research groups are moving in this direction \cite{denicol,jaki,bass,pratt,schenke,bozek1}.

These models are all based on the reduction, either exact or approximate, of the problem to a two-dimensional system \cite{bjorken}, based on the symmetry of boost-invariance.  Essentially, the system at mid-rapidity is assumed to have {\em as an initial condition} a longitudinal flow that is Hubble-like in the beam direction (usually associated with the $z$ coordinate) {\em only}.    This means that, initially, 
\begin{equation}
\label{boost1}
v_z = \frac{z}{t} \eqcomma y_{s}=y_{f}=\ave{y}_{p},
\end{equation}
where $y_{s}$ and $y_f$ are respectively the spacetime and flow rapidities
\[\
y_{s} = \frac{1}{2} \ln\left(\frac{t+z}{t-z} \right)
\eqcomma y_{f} = \frac{1}{2} \ln\left(\frac{1+v_z}{1-v_z} \right),
\]
\[\
\ave{y}_{p} = \frac{1}{2} \ave{\ln \left(\frac{E + p_z}{E-p_z}  \right)} \simeq \frac{1}{2} \ln \cot \frac{p_z}{|p|}
\]
with $y_p$ being usually referred to as pseudo-rapidity.
A further simplification comes from assuming that all initial dynamics does not depend on $y$
\begin{equation}
\label{boost2}
\frac{d}{dy}\frac{dN}{dy}=0 \eqcomma \frac{dv_T}{dy}=0
\end{equation}
or, equivalently not on $t,z$ separately, but just on 
\begin{equation}
\label{propertime}
\tau = \sqrt{t^2-z^2}
\end{equation}
 (evolved from an initial time $\tau_0$) and transverse degrees of freedom.  An initial condition that respects \refeq{boost1} but not \refeq{boost2} will slowly degrade the constraints of \refeq{boost1}, as shown in \cite{mish,bozek,kodama}.
2+1 dimensional codes typically assume both \refeq{boost1} and \refeq{boost2}.  3+1 dimensional codes can relax either of these assumptions 
but will yield results at mid-rapidity approximately identical to
   \refeq{boost1} and \refeq{boost2}, {\em if boost-invariance is
   assumed as an initial condition}.

Initially, a different model has originally been advocated as the obvious initial state for the hydrodynamic evolution of the fluid: Landau hydrodynamics
\cite{landau,landau2,landau3,landau4}.
In this picture, the energy that forms the bulk of the expanding fireball ``stops'' at midrapidity at time zero (in the collider frame).    The initial distribution of matter is therefore a ``pancake'', of thickness $2\Delta$ related to the boosted charge radius $R$ of the nuclei with nucleon mass $m_{N}$ at center of mass energy of $\sqrt{s_{NN}}$ where
\begin{equation}
\label{deltadef}
\Delta \rightarrow \Delta_{lim} \simeq \frac{R}{\gamma} = \frac{Rm_{N}}{(\sqrt{s_{NN}}/2)} .
\end{equation}
In a more general implementation, $\Delta$ need not be defined by \refeq{deltadef} and can be a free parameter, reflecting the spread in configuration space of low $x$ gluons.  The initial Landau condition is defined by the assumption that 
the initial ``pancake'' has no existing longitudinal flow at all, unless there are initial inhomogeneities which lead to a net momentum in local transverse space. (This is known as the ``firestreak model'' \cite{firestreak,firestreak2}).
Boost invariance is badly broken at the beginning of the fireball evolution and such a pancake has very little in common with the scenario used in \cite{bjorken,bozek}.    One can consider Bjorken and Landau as two extremes:  In the Bjorken scenario, the nuclei originally pass through each other with minimal reinteraction and strings that stretch between colliding gluons arise in parallel to other strings.   In the Landau scenario, they ``stick together'' or at least leave some energy in the middle.

While constructing a coordinate system around a physical symmetry is highly desirable, a physics justification would be needed for the approximation of \refeq{boost1}.   A direct measurement of $dN/dy$ is inconclusive.
On the one hand, the Landau model fits a Gaussian well at all energies, with universal limiting fragmentation, as expected in \cite{landau,landau2,landau3};  moreover, strong violations of boost-invariance considerably lessen the HBT puzzle \cite{gyulhbt}. However, the {\em  multiplicity} dependence on $\sqrt{s_{NN}}$ is not exactly that predicted in \cite{landau,landau2,landau3}.     This by itself does not rule out the Landau scenario, as it can be accounted for by treating the initial thickness evolution with $\sqrt{s_{NN}}$ as a free parameter, as done in the Bjorken scenario.     

\begin{figure}[h]
\epsfig{width=0.35\textwidth,figure=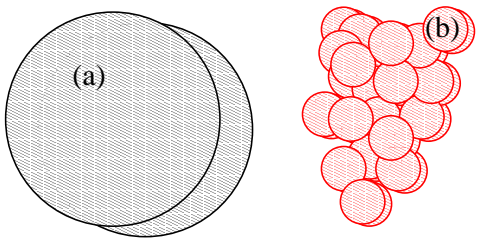}
\caption{\label{pancakes} 
 Landau initial condition (panel (a)), and actual Glauber  
initial conditions (panel (b)) for a typical event.
}
\end{figure}
There are two main arguments one can give for the Bjorken limit being more appropriate:  The first one is that the perturbative partonic picture of the system \cite{feynman,bgk} makes this initial condition natural.  However, even in the weakly-coupled limit,  low $x$ partons could lead to a breakdown of  \refeq{boost2} (see for example, \cite{busza,dpm}).
Moreover, if the initial state is {\em strongly coupled} from the beginning, one could indeed expect that it would appear much more Landau-like \cite{kovland,romland} than Bjorken-like, although the degree of stopping might also depend strongly on energy and system size \cite{wilke1,wilke2}.   Stopping is therefore not determined {\it a priori}, as 
 the interaction strength at the beginning of the system's evolution is currently a controversial topic.

The second reason is that, for mid-rapidity data, it is widely believed that the distinction between Bjorken and Landau evolution is irrelevant.
As is clear from \cite{landau}, Landau evolution converges to Bjorken evolution after some sufficient time.
The reason for this behavior is that longitudinal flow forms on the scale of $\sim \Delta /c_s$, while transverse flow forms on a much larger scale $\sim R/c_s$ where $c_s$ is the speed of sound.
Hence, since $\Delta \ll R$ for $\sqrt{s_{NN}} \gg 1$ GeV, initially the system can be considered, as indeed it is in \cite{landau}, to be a purely 1D expanding ``sharp step.''
As again shown in \cite{landau,landau3,landau4}, the long-term longitudinal evolution of such a system at mid-rapidity is indistinguishable from that of \cite{bjorken}.    Hence, boost-invariant hydrodynamics can be safely used even if, at the very initial stage \cite{bozek}, the system is very far from boost invariance.
Landau evolution at mid-rapidity can be treated as Bjorken with $\tau_0 \sim 1\phantom{A}\mathrm{ fm}\times \mathrm{GeV}/\sqrt{s_{NN}}$.   Perhaps, this scaling will give unrealistically low initial proper times at the LHC, but since boost-invariant simulations are only weakly sensitive to time \cite{jaki}, this might not be a fatal issue.
\begin{figure}[h]
\epsfig{width=0.44\textwidth,figure=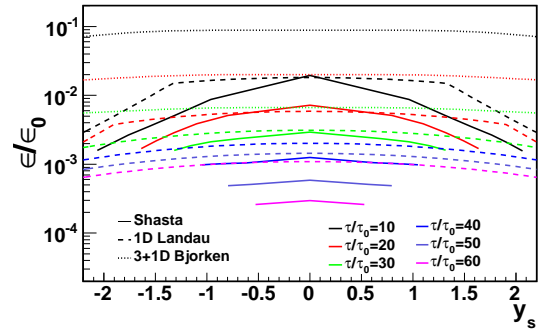}
\caption{\label{evolution} 
(color online) Ratio of energy density at the indicated time to the initial energy
   density as a function of rapidity.  
The dashed lines show the
   analytical solution in (1+1)D \cite{wonganal} while the solid lines
   show our numerical calculation in (3+1)D, with Landau initial
   conditions, including net momentum, transverse expansion and
   inhomogeneities.  The results from a (3+1)D hydrodynamics where
   Bjorken-type boost-invariant longitudinal flow is set as an initial
   condition \cite{hirano1,hirano2} are also shown as dotted lines.
Time is normalized to units of $\Delta$ in the Landau picture and $\tau_0$ in the Bjorken picture.
}
\end{figure}
This idea, however, has two flaws:  First of all, for a non-central collision, where anisotropic flow is most expected, locality and  longitudinal momentum conservation imply that the system develops an additional initial longitudinal momentum imbalance, with extra longitudinal momentum due to the local (in transverse space) imbalance between the target and projectile $\rho_{part}^{P,T}(x_T)=d^2 N_{part}^{P,T}/dx_T^2$ transverse participant density.   Momentum conservation and the Landau condition (no transparency) constrain the initial $\gamma_z v_z$ to
\begin{equation}
\label{torque}
\frac{ v_z(x_T)}{\sqrt{1-v_z^2(x_T)}} = \frac{\rho_{part}^P(x_T) - \rho_{part}^T(x_T)}{\rho_{part}^P(x_T) + \rho_{part}^T(x_T) }  K
\end{equation}
$K$ here is a free parameter, but it is clear that 
$K=\sqrt{s_{NN}}/m_N$ when $\Delta=\Delta_{lim}$ in \refeq{deltadef}.  In general, $\Delta > \Delta_{\lim}$ reflects a picture where the partons carrying the dominant fraction of the nucleon's energy are parametrically much softer than the nucleon.    This is equivalent to the ``wee parton" picture, and implies they also carry less momentum.   Assuming a linear dependence, the net momentum in an off-central collision is related to $\Delta$ by
\begin{equation}
\label{kdef}
K \simeq \frac{\sqrt{s_{NN}}}{m_N} \frac{\Delta_{lim}}{\Delta}= \frac{2R}{\Delta}.
\end{equation}
This initial flow is trivially not boost invariant and it is not clear it disappears at {\em any} finite time for a general system evolving from a Landau initial condition.

Additionally, the ``Landau$\rightarrow$Bjorken''  reasoning assumes that the longitudinal timescale is much larger than the transverse one.   This is certainly true if the transverse scale is given as a radius of a homogeneous ``pancake'' of radius $R$ given by an {\em average} of many events as in Fig. ~\ref{pancakes} (a).
It is however less clear that such a hierarchy holds for a {\em typical} event as in Fig. \ref{pancakes}(b).
The inclusion of subnucleonic strong QCD fields \cite{colemansmith,schenke}
 make this hierarchy even more dubious as the events with the strongest anisotropic coefficients would also have the most prominent ``hotspots.''
Potentially, this effect makes the boost-invariant picture irrelevant {\em even for late-time hydrodynamics}: The more homogeneous regions will be more similar in their longitudinal expansion to \cite{bjorken}, while the more inhomogeneous regions would, on their own, evolve to a {\em 3D Hubble} expansion \cite{saito}.
The interplay between regions of different symmetry, and local instabilities \cite{saito,usstable} makes any symmetry dubious.

\begin{figure}[h]
\epsfig{width=0.5\textwidth,figure=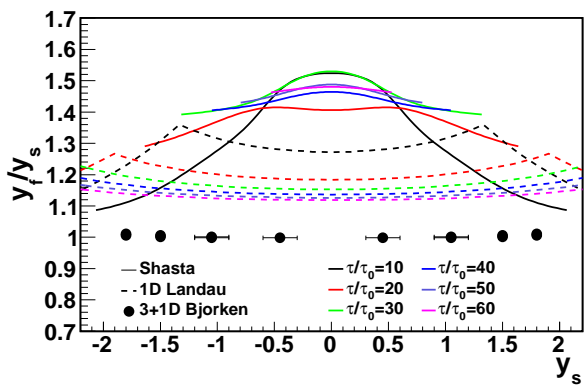}
\caption{\label{figratio} 
(color online) The $y_{f}/y_{s}$ ratio (solid lines) as a function of proper time
   and rapidity, averaged over all events, at $r_\perp=0$ in (3+1)D
   numerical hydrodynamics with the Landau initial condition.  The
   analytical (1+1)D solution \cite{wonganal} is also plotted (dashed
   lines) The results of a (3+1)D calculation with Bjorken
   boost-invariance assumed as an initial condition are also shown
   \cite{hirano1,hirano2} as solid points. Time is normalized to $\Delta$ in the Landau picture and $\tau_0$ in the Bjorken picture}
\end{figure}

\begin{figure*}[t]
\epsfig{width=0.84\textwidth,figure=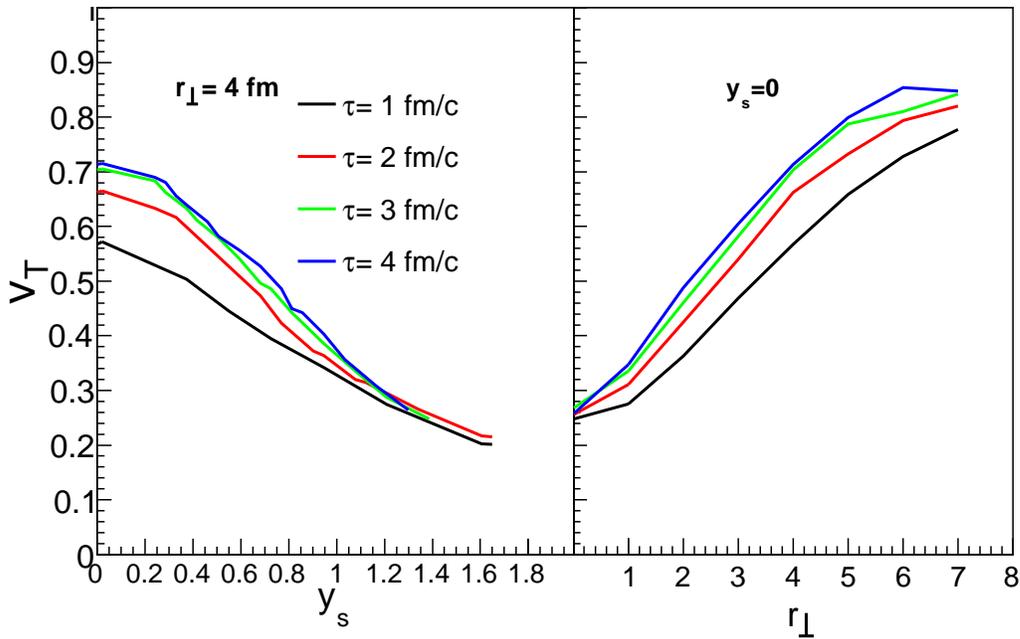}
\caption{\label{transverse_vel} 
(color online) Transverse velocity as a function of rapidity (left panel) and $r_\perp$ (right panel) at several longitudinal proper times, averaged over all events.   
}
\end{figure*}

To investigate these effects further, one needs to perform (3+1)D calculations starting from Landau initial conditions and transverse inhomogeneities.  
In this work, we use an event-by-event Glauber model to generate initial-state transverse energy distributions, with the longitudinal density distribution being given by a Landau profile. 

The Glauber Monte Carlo description of two colliding $Au^{197}$ nuclei at 200 GeV was used to generate the initial condition relevant to RHIC. Nucleons were distributed as per a Wood-Saxon distribution with radius 6.38 fm, and diffuseness 0.535 fm. The impact parameters were simulated randomly following a distribution of $d\sigma/db= 2\pi b$. 
The nucleons were assumed to have no hard-core and the condition for nucleon-nucleon collision is that the inter-nucleon distance $d$ should satisfy $\pi d^2<\sigma_{NN}$, where $\sigma_{NN}=42$ mb is the nucleon-nucleon cross section.

We then use the CL-SHASTA code developed in \cite{clshasta} to evolve this configuration according to ideal hydrodynamics,
$\partial_\mu T^{\mu \nu}=0$ with
\begin{eqnarray}
T_{\mu \nu}=(\rho+p)u_\mu u_\nu +pg_{\mu \nu}
\end{eqnarray}
\begin{eqnarray}
u_\mu = \frac{1}{\sqrt{1-v_z^2 - v_T^2}} \left( 1 , v_T \sin(\theta),v_T \cos(\theta),v_z \right)
\end{eqnarray}
where $v_z = \tanh y_L$.
 With an ideal gas equation of state, $p=\rho/3,c_s=1/\sqrt{3}$, and $\Delta=0.1$ fm, and longitudinal flow given by \refeq{torque}.
Our results do not qualitatively change if the longitudinal thickness if changed by $\order{50-100\%}$.    

The high-statistics (3+1)D calculations were performed at the Oak Ridge National Laboratory using the code in \cite{clshasta}.
The availability of the TITAN supercomputer facility at Oak Ridge Leadership Computing Facility allows us to collect an ensemble of these numerically intensive calculations which is large enough to explore event-by-event correlations. For relativistic hydrodynamical calculations, the (3+1)D Sharp and Smooth Transport Algorithm (SHASTA) was recently completely rewritten using the OpenCL computational framework to work on accelerators like Graphic Processing Units (GPUs). Parallelized algorithm kernels written in OpenCL run on GPUs with concurrent execution of thousands of streams. For this letter, adjustments were made for optimal use of the powerful NVIDIA GPUs of the TITAN supercomputer. Using redesigned algorithms and harnessing the processing power of GPUs, the hydrodynamical calculations have been accelerated by a factor $\sim$100x for a given node, scaled to a large number of Titan nodes. This allowed us to accumulate a large ensemble of event-by-event statistics with unprecedented efficiency for relativistic hydrodynamical simulations.
In order to organize the hydrodynamic expansion into thousands of execution streams, the problem is reduced by {\it domain decomposition}. This leads to a grid structure in the spatial dimensions where the grid elements are still connected but can be processed separately. The grid size depends on hardware and algorithm type. The current implementation of the grid includes 8 million grid cells, which covers $\pm 10$ fm in each spatial dimension. Each grid cell holds the physical properties in that spatial region and one kernel per physical quantity is used to modify them accordingly through out the expansion.

After simulating 10000 events for a given configuration and initial conditions, each of which evolves the millions of grid cells over 300 small time steps in the lab frame covering an expansion until 10 fm/c, we divide them into spacetime rapidity slices.   We also compare with a (3+1)D  hydrodynamic code where boost-invariance was initially assumed, \cite{hirano1,hirano2}\footnote{The results are publically available at\\
http://tkynt2.phys.s.u-tokyo.ac.jp/$\sim$hirano/parevo/parevo.html\\
or via the TECHQM webpage\\
https://wiki.bnl.gov/TECHQM/} and \cite{shen1,shen2} where boost-invariance has been enforced as an initial condition and we concentrate on the early dynamics where the hadron gas contribution is negligible. 

The energy density evolution as a function of spacetime rapidity is shown in Fig. \ref{evolution}, which follows the trend in \cite{landau} to $\sim$20\% precision, as expected from correction due to transverse and elliptic flow.   
Hence, Fig. \ref{evolution} show a decreasing bump in $\epsilon$ (which correlates with transverse multiplicity$dN/dy \sim S \tau_0 \epsilon^{3/4}$ and transverse energy  $dE_T/dy \sim S \tau_0 e$ in the Bjorken picture \cite{bjorken}.  Here $S \sim R^2 \sim N_{part}^{2/3}$ is the transverse overlap area), making these similar to boost-invariant results \cite{bjorken}.  While comparing with  the boost-invariant calculation from \cite{hirano1,hirano2} should be done with care as the physical meaning of $\Delta$ and $\tau_0$ are different, such a comparison confirms that for realistic time-scales the evolution
in the two limits is significantly different.
  Self-quenching variables ($v_2$, and to a lesser extent the average transverse momentum $\ave{p_T}$), however, will be sensitive to such differences independently of freezeout.

We then calculate the longitudinal flow rapidity $y_f$ as well as the transverse flow for each slice of rapidity, averaged over the entire transverse volume, to explore boost invariance.   
Fig. \ref{figratio} shows the ratio $y_{f}/y_{s}$ as a function of $y_s$ at various relevant times in the evolution.   If the system were exactly boost invariant, $y_{f}/y_{s}$ would be strictly unity.    Moreover, as Fig. \ref{figratio} also shows $y_f/y_s$ averaged over both transverse volume and proper time in a (3+1)D evolution  where boost-invariance is set as an initial condition \cite{hirano1,hirano2} (3+1)D dynamics acts as a very small correction to the longitudinal flow over the realistic timescale of the evolution.   This shows that when Bjorken flow is added as an initial condition, (3+1) dimensional hydrodynamics will be a small correction over 2+1 dimensional hydrodynamics.

In the Landau limit the ratio does evolve towards unity as the system cools; however, it would be a gross oversimplification to treat the ratio as a constant or unity, even at significantly later times.   At freezeout provided initial temperature $\simeq 300$ MeV, we predict $y_{f}/y_{s}$ to be above unity by about $40\%$ around midrapidity. At earlier stages, relevant for the formation of transverse and elliptic flow ($t\sim \epsilon_r R/c_s$ where $\epsilon_r$ is the eccentricity), these corrections are of order $50\%$.
For comparison, we superimpose the same distributions for the 1D expansion calculated in an analytical work \cite{wonganal}.   It can be seen that, unlike what was presumed in \cite{landau2,landau3}, transverse expansion and local dynamics make a qualitative, and not just a quantitative effect: deviation from boost invariance oscillates and stays nearly constant rather than decreases in time when transverse expansion and anisotropies are taken into account.   

This discrepancy is directly confirmed in Fig. \ref{transverse_vel} which shows that the transverse velocity as a function of the spacetime rapidity significantly violates \refeq{boost2}, with an apparent decrease of $v_T$ as the system expands.  This apparently counter-intuitive behavior can be explained by the fact that when the rarefaction wave traverses the system size, the outer-going shock could well experience a negative gradient at the point of maximum density (the density in front of the wave's peak, determined by the shock wavefront, is {\em higher} than the density behind it, given by the rarefaction wave).   Longitudinal expansion weakens this effect by depleting density in all of transverse space at the same time, but, as our simulation shows, in the Landau limit the full 3D flow development could be non-monotonic for part of the evolution.  
Note that Fig. \ref{transverse_vel} also shows that, while transverse velocity increases with $r_\perp$ as usually predicted, the presence of hotspots may make the average magnitude of $v_T$ non-zero at $r_\perp=0$, with a rapidity dependence which follows the  Gaussian profile characterising the event (its direction of course averages to zero, but it is non-zero in a typical event).

\begin{figure*}[t]
\epsfig{width=0.83\textwidth,figure=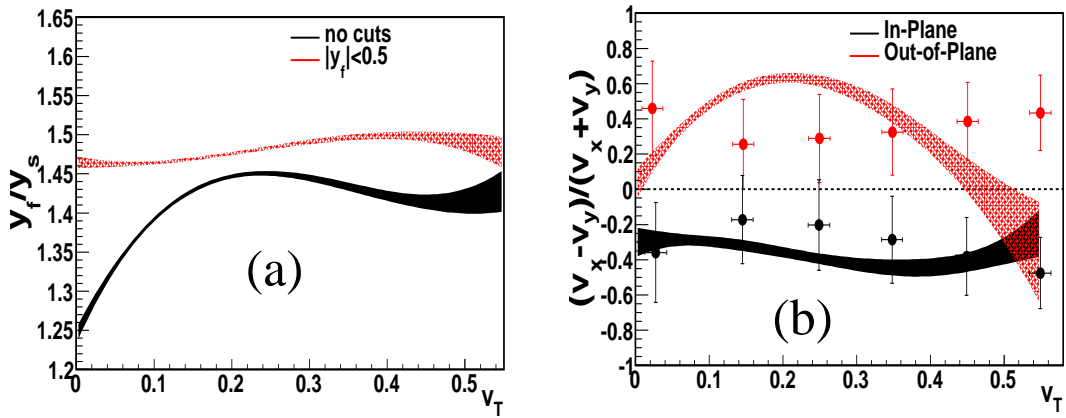}
\caption{\label{corr} 
(color online) Panel (a) $y_f/y_s$ as a function of mean transverse velocity for $\mod{y_f}<0.5$ (hashed red) and with no restriction on $y_f$ (solid black).
Panel (b) Distribution of transverse flow anisotropy versus mean total transverse velocity for in-plane (solid black) and out-of-plane (hashed red) flow.
Both are averaged over fireball volume and event sample, with the bands representing the variance over the average across events.   The results for event-by-event (2+1)D simulation including initial inhomogeneities \cite{shen1,shen2} are also shown as points.
}
\end{figure*}

The relevance of this dynamics for transverse degrees of freedom is further elucidated in Fig. \ref{corr}, which shows the dependence of $y_f/y_s$, an indicator of the degree of violation of boost invariance, on transverse flow.
Thus, if the Landau initial condition is more appropriate, transverse flow and its azimuthal anisotropies form, to a certain extent, in strongly non-boost invariant regions. This is readily understood, as such regions are precisely the places where transverse gradients are larger w.r.t. longitudinal ones.    Hotspots can also have a non-zero longitudinal momentum and vorticity \cite{betz} (the ``firestreak''), further invalidating 
local boost-invariance.
As Fig. \ref{corr} panel (a) however shows, this result somewhat depends on the rapidity region being explored.
A restriction in flow rapidity, approximately tracking the pseudorapidity, will ensure $v_T$ is independent of the degree of boost invariance.   Such a cut, however, does nothing to make the evolution examined more boost-invariant, since $y_f/y_s$ remains very well away from unity.

Fig. \ref{corr} panel (b) shows the anisotropy of the in-plane and out of plane flow, as a function of transverse flow. The combination of the results of Figs. \ref{transverse_vel}, \ref{corr} indicates that dynamics relevant for transverse and anisotropic flow significantly violates boost invariance if Landau initial conditions are assumed.     This is confirmed by comparing our results to the
flow profile of \cite{shen1,shen2}: The correlation between anisotropy and flow is significantly weaker, and qualitatively different-looking in the Landau than in the Bjorken limit throghout the evolution of the fireball: Whereas in the Landau limit flow eccentricity is maximized in the middle of the fireball, in the Bjorken limit it is maximized at the edges.    This is because in the Bjorken limit there is no interplay between transverse and longitudinal flow, whereas
in the Landau limit the longitudinal ``twist'' in the $z-x$ and $z-y$ direction is developed contemporarily with the $x-y$ flow.
Hence, a value of $\eta/s$ w.r.t. that fitted in papers where initial longitudinal flow was assumed \cite{bass,schenke} will most likely be required to fit flow harmonic data with Landau rather than Bjorken initial longitudinal conditions.
By dimensional analysis, this difference should be parametrically comparable to the deviation between the Landau and Bjorken model shown in Fig. \ref{corr} (b).

Indeed, the main shortcoming of this analysis is that the hydrodynamics was assumed to be ideal.
However, it should be noted that viscosity is sensitive to differences between $y_{s}$ and $y_{f}$ examined here in a way which may be different from the intuition from boost-invariant hydrodynamics.   
Viscosity, shear and bulk, transforms gradients into heat.
This suppresses the local structure of flow, but it also creates extra pressure that enhances flow in all directions.   It has been recently realized \cite{jaki} (in a model incorporating bulk viscosity, for which the first effect is reduced)  the second effect's contribution to $v_n$ 
 can be {\em positive}, since heat creation enhances local pressure gradients, thereby boosting transverse expansion, which enhances all remaining flow structure, and this can overpower the direct degradation of flow gradients by viscosity.

Since $v_n$ is gradient projection in a purely transverse direction, for {\em longitudinal} gradients this  degradation is minimized and hence viscous heating could overpower it. 
For boost-invariant hydrodynamics longitudinal gradients are fixed at
$\sim \tau^{-1}$, and hence direct suppression of $v_n$ by viscosity overpowers viscous heating, as amply confirmed by numerical simulations  \cite{bass,pratt,schenke,bozek1}.
As our work shows, in Landau hydrodynamics the longitudinal gradient is much greater than $\tau^{-1}$ even at mid-rapidity.   The extra boost in the gradient can slow down cooling without affecting azimuthal gradients.    Thus, if initial conditions are more Landau-like, {\em shear} viscosity could be significantly higher than what is inferred by boost-invariant calculations, and could even be {\em correlated} rather than {\em anti-correlated} with initial eccentricity.

\begin{figure}[h]
\epsfig{width=0.55\textwidth,figure=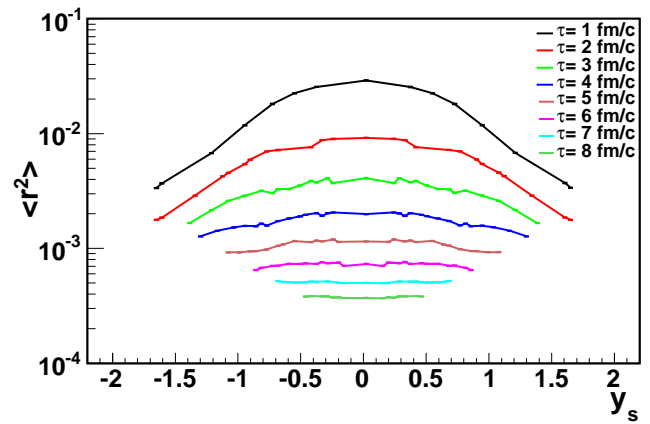}
\caption{\label{fighbt} 
The event-average transverse size of the system as a function of $\tau$ and rapidity, for all events averaged over centrality. Overall normalization is arbitrary up to a factor constant in time, rapidity, and transverse shape.
}
\end{figure}

The viability of the computations performed here depends, of course, in the longitudinal structure of the event really being close to the Landau limit. 
Because we do not know this from first principles, and given the many undetermined parameters in a typical hydrodynamic simulation, we suggest that experimental tests specifically probing  boost invariance should be performed.    
It is intuitively clear that in the Bjorken solution the transverse size of the system, along with other parameters, does not vary with rapidity.  It is equally intuitively clear that the strong dependence of flow with rapidity produces a strong rapidity dependence of size at late times.   Fig. \ref{fighbt} confirms this, where the average $\ave{r^2}$ integrated over the transverse radius is shown as a function of rapidity.   As can be seen, it approximately follows the Gaussian structure of the transverse momentum characteristic of Landau hydrodynamics \cite{landau,landau2,landau3}, varying over orders of magnitude in the fragmentation region.   In the Bjorken picture, such wide variation is excluded since the transverse size is bounded by the {\em initial} transverse size, $\sim N_{part}^{2/3}$ at all rapidities.

This quantity, in the Gaussian approximation, is related to the HBT variable $R_{side}$ \cite{mikloshbt}.   This relationship is not straightforward, since $R_{side}(K)$ is defined in terms of a momentum pair $K$, and will yield, approximately \cite{mikloshbt} the ``homogeneity'' region, the region from which ``typical'' particles of momentum $K$ are emitted (this relation comes out explicitly out of integrating the emission function).   However, experimental data shows \cite{kisiel} that this subtlety does not change the geometric scaling of all HBT radii, on which our proposal underlies.    
Furthermore comparing Fig. \ref{fighbt} with Fig, \ref{transverse_vel} it becomes clear that the observed $R_{side}$ will be {\em steeper} than $dN/dy$ because away from mid-rapidity the emission volume is smaller and less out-flowing.   Thus, in those regions particles will be emitted from a smaller surface and an earlier time, less affected by expansion.
In the Bjorken picture, where the initial state is a "cylinder" in rapdity, the rapidity independence of the system size should not produce such a steep decrease even if the initial density has some rapidity dependence.    
Therefore, a steeply falling experimentally measured HBT $R_{side}$ for pairs in different rapidity bins would be good evidence of a Landau-like initial condition for hydrodynamic evolution.

\begin{figure*}[t]
\epsfig{width=0.85\textwidth,figure=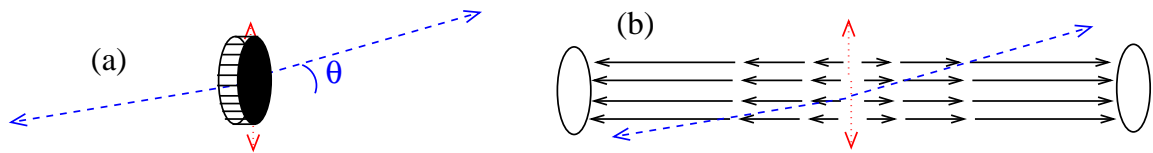}
\caption{\label{figjet} 
A schematic representation of tomography in a Landau (panel (a))  vs Bjorken collision (panel (b)) for pairs of jets produced at higher rapidity (dashed lines with arrows) and mid-rapidity (dotted lines with arrows).  While the quantitative result depends on the details of the quenching model (see for example \cite{abc}), it is clear $R_{AA}$ will rapidly increase with rapidity in the Landau limit.
}
\end{figure*}

In the same way, the shorter longitudinal size of the Landau ``pancake'' in spacetime would mean that in-medium energy absorption  for a fast parton ("tomographic energy loss") for higher rapidity will be significantly weaker than in the purely transverse direction (Fig. \ref{figjet} panel (a)), since the initial size in the longitudinal direction in the Landau limit will be much smaller ($\sim \Delta/\sin\theta$) than in the Bjorken limit ($\sim R/\cos(\theta)$, Fig. \ref{figjet} panel (b)), and since jets will generally traverse the system faster than the build-up of longitudinal flow.     Note that this distinction is sensitive to precisely the physical difference of Landau and Bjorken: In the Bjorken case the projectile and target collide transparently and continue moving at the speed of light (faster than the fast parton's speed), while in the Landau case longitudinal motion stops until hydrodynamics sets in (parametrically slower than the fast parton's speed).

The decrease in longitudinal size of course is balanced by the higher initial density, but away from the weakly coupled Bethe-Heitler limit, size and density do not compensate  \cite{jetquenching}.
For instance, in the LPM limit the total energy lost by the parton traversing a medium of length $L$ is $\Delta E \sim \rho L^2$, while if theories with gravity duals describe jet-medium interaction, the energy lost by the parton $\Delta E \sim \rho L^{m\geq 2}$ ( \cite{abc} and references therein).
Following the calculation of $R_{AA}$ in \cite{Jia}, where this exponent $m$ is kept arbitrary
\begin{equation}
R_{AA} \simeq \ave{\exp \left[- \kappa \int dl l^{m-1} \rho \left( x_0 + \hat{n} l \right)  \right]}
\end{equation}
where $\kappa$ is a constant and $\ave{\cal{A}}$ integrates $\cal{A}$ over all $x_0,\hat{n}$ and events.
We can use simple geometrical scaling from Fig. \ref{figjet} to approximate the trigger particle's rapidity $y$ by the pseudo-rapidity
\begin{equation}
y \simeq     -   \ln \tan \left( \frac{\theta}{2} \right)
\end{equation}
Assuming fast partons are produced at $y_s=0$ (at the initial collision), and uniformly in net Bjorken $x$,
we infer that, if the jet leaves the system before significant flow develops, for $y \gg 0$ and in terms of $\kappa' = \kappa \ave{\rho \Delta^{m}}$.  In the limit $y\gg 0,\Delta \ll R$
\begin{equation}
\label{raaland}
R_{AA}(y) \sim   \exp \left[ - \kappa' \exp \left[ -m y \right]  \right]
\end{equation}
No doubt this estimate is extremely rough, and a more quantitative estimate is the subject of a subsequent work, but, unless the bulk of jet energy loss is due to {\em non}-tomographic effects (such as initial \cite{kopel} and fragmentation \cite{colorflows} effects) or jet energy loss is {\em not} approximately collinear (as is generally believed), we can expect
that the jet suppression parameter $R_{AA}(p_T,y)$ \cite{jetquenching} to rise steeply with the rapidity of the trigger particle $y$.   In contrast, since in the Bjorken limit jets are still produced during the initial hard scattering $y_s=0$ for {\em all} $y$, the corresponding quantity to  Eq. \ref{raaland} is
$
R_{AA}' \sim \exp \left[ - \kappa' ]    \right]
$ independently of $y$, 
since the parton keeps traversing the medium even at $y\gg 1$.     In this limit, $R_{AA}'$ is not a good approximation as longitudinal expansion is neglected \cite{abc}, but one expects that
tomographic energy loss should {\em not} decrease in rapidity even for very high rapidities, and may in fact increase if the $m$ parameter \cite{Jia} is large enough for the extra path (Fig. \ref{figjet} panel (b)) to compensate for decreased in-medium parton density.
Thus, jet energy loss dependent on rapidity could be a decisive and direct test of boost invariance.
While experimental results do tend to favor a Bjorken picture rather than the picture examined in this paper \cite{raabrahms,raaatlas}, a systematic study relating $dN/dy$ to $R_{AA}$ in rapidity, as well as a quantitative calculation of $R_{AA}$ in both limits, is necessary for a definite conclusion.

   Observables such as the correlator studied in \cite{gavin} or polarization \cite{betz} could give further tests.
 
In conclusion, we have shown that, provided the system is Landau-like in its initial stages, it will not, as commonly expected, evolve to a Bjorken-like stage within realistic timescales.
Furthermore, the deviation from boost invariance is directly correlated with the development of transverse and elliptic flow, the characteristic signatures used to demonstrate and quantitatively study the hydrodynamics of the quark-gluon plasma.   In view of these results, the transport properties of the medium created in heavy ion collisions could be considerably different from those usually assumed.

This research used resources
of the Oak Ridge Leadership Computing Facility at the Oak Ridge National
Laboratory, which is supported by the Office of Science of the U.S.
Department of Energy under Contract No. DE-AC05-00OR22725. 
GT also acknowledges support from FAPESP proc. 2014/13120-7
We wish to thank Sean Gavin and Peter Steinberg for useful discussions.


\begin{thebibliography}{19}
\bibitem{shen2} We would like to thank Chun Shen for modifying the code to make the data presented available, and for assistance with installing the code.
\end{thebibliography}
\end{document}